\documentclass[final,3p,times,twocolumn]{elsarticle}

\usepackage{amssymb}
\usepackage{amsthm}
\usepackage{amsmath,bm,amssymb}
\usepackage{url}

\journal{Computer Physics Communications}

\begin{document}

\begin{frontmatter}



\title{Classical Radiation Reaction in Particle-In-Cell Simulations}


\author[label1]{M. Vranic}
 \ead{marija.vranic@ist.utl.pt}
\author[label1]{J. L. Martins}
\author[label1,label2]{R. A. Fonseca}
\author[label1]{L. O. Silva}
\address[label1]{GoLP/Instituto de Plasmas e Fus\~ao Nuclear, Instituto Superior T\'ecnico, Universidade de Lisboa, Lisbon, Portugal}
\address[label2]{DCTI/ISCTE - Instituto Universit\'ario de Lisboa, 1649-026 Lisboa, Portugal}

\begin{abstract}
Under the presence of ultra high intensity lasers or other intense electromagnetic fields the motion of particles in the ultrarelativistic regime can be severely affected by radiation reaction. The standard particle-in-cell (PIC) algorithms do not include radiation reaction effects. Even though this is a well known mechanism, there is not yet a definite algorithm nor a standard technique to include radiation reaction in PIC codes. We have compared several models for the calculation of the radiation reaction force, with the goal of implementing an algorithm for classical radiation reaction in the Osiris framework, a state-of-the-art PIC code. The results of the different models are compared with standard analytical results, and the relevance/advantages of each model are discussed. Numerical issues relevant to PIC codes such as resolution requirements, application of radiation reaction to macro particles and computational cost are also addressed. For parameters of interest where the classical description of the electron motion is applicable, all the models considered are shown to give comparable results. The Landau and Lifshitz reduced model is chosen for implementation as one of the candidates with the minimal overhead and no additional memory requirements. 
\end{abstract}

\begin{keyword}
radiation reaction \sep particle-in-cell \sep laser-matter interactions \sep relativistic electron motion


\end{keyword}

\end{frontmatter}

\section{Introduction}


The next generation of high-power lasers is going to reach intensities that will
open new doors for exploring a wide range of physical problems with 
an even wider range of applications. The ELI project \cite{ELI} is expected to reach laser intensities 
several orders of magnitude higher than those available today.   At intensities $I \sim 10^{23}-10^{24} \ \mathrm{W/cm^2}$ one can 
expect electron-positron pair production \cite{model1bell, Bulanov_pairsvaccuum, Ridgers_solid, Elkina_rot}. In astrophysics these intensities are relevant for the study of pulsars, blazars, and gamma-ray bursts \cite{model2noguchi}. 
High intensity laser-mater interactions can also produce proton and heavy ion beams \cite{model2zhidkov, application1, application2, ion_acc_rradd} that are of great significance for many applications, the most 
important being cancer treatment. 

At high intensities particle acceleration can be severely limited by the radiation reaction associated with the energy loss via radiation emission \cite{limits}. This is important whenever the radiated energy is comparable to the total particle energy. The threshold electromagnetic field intensities to drive these effects vary in the available literature, but they are usually in the range $10^{22}-10^{25} \ \mathrm{ W / cm^2}$. Many authors have discussed the classical radiation reaction in pursuit of a proper analytical description \cite{rorlich_1, Piazza_solutionLL, counter, Self_force_Derivation, Bulanov_LLLAD, Keita_neweq, spohn, Fedotov_ponder, Dino_RR, Noble_2013_rrmodel} or experimental signatures  \cite{RRinCascade_classical, Rohrlich_derivation, Tamburini3d, Bulanov_chi, Tikhonchuk, Bulanov_review, ThomasRR, emittance_decrease, Pukhov_rrtrapp, Harvey_inpulse, Capdssus_ionwithrr}. There is also a rising interest in the effect of the radiation reaction on particle dynamics in astrophysical phenomena \cite{JaroschekHoshino, theory_rrinmr, Angelo_astro, Cerruti_apj_2013}. In order to perform reliable particle-in-cell (PIC) simulations in the classical radiation-dominated regime, radiation reaction (RR) must be included in the equations of motion for the particles. Though several models for classical radiation reaction have been proposed in the literature, there is not a definite standard choice to apply in PIC codes. In this paper we compare several different models in order to find the most appropriate one for implementation in OSIRIS  \cite{OSIRIS, Fonseca_scaling}. The chosen model should capture all the relevant physics in the scenarios we are aiming to explore at the lowest possible cost in performance. We test each of them with well studied examples of particle motion in electromagnetic fields where the trajectory can be analytically expressed and estimates for the radiated power/energy can be obtained. Our analysis shows that even though there are conceptual differences between the models considered, in the parameters of interest that could be tested with near-future laser technology all the models give the same description of particle motion. To observe differences, one would need to impose a linear acceleration so high that an extreme electric field required to produce it would render a classical description of an electron trajectory inapplicable. The choice can then be made on the basis of the computational overhead and with this aim the additional computational cost that the implementation of each model introduces is presented. Among the models with the lowest computational requirements, we opted to introduce Landau and Lifshitz reduced model in OSIRIS framework. Specific questions associated with the interpretation of classical RR in PIC simulations are also addressed. 

This paper is structured as follows. In section 2 we introduce the radiation reaction models and deal with macro particle interpretation. In section 3, the behaviour of all the models is investigated in the standard cases of synchrotron radiation and bremsstrahlung. Section 4 underlines the difference between the results with and without radiation reaction for an electron in a laser pulse field and gives an estimate of the threshold for the detectable radiation reaction.  The issue of the optimal temporal resolution is addressed in section 5, and  section 6 contains estimates for the computational overhead for each model. Finally, in section 7 we state the conclusions. 

\section{Radiation reaction models}

The charged particle motion with radiation reaction is expressed by the Lorentz-Abraham-Dirac (LAD) equation   \cite{Jackson}:
\begin{align}\label{LAD}
\frac{dp_\mu}{d\tau}&=F_\mu^{EXT}+F_\mu^{RR}\quad
\mathrm{where} \nonumber \\
F_\mu^{RR}&=\frac{2e^2}{3mc^3}\left(\frac{d^2p_\mu}{d\tau^2}+\frac{p_\mu}{m^2c^2}\left(\frac{dp_\nu}{d\tau}\frac{dp^\nu}{d\tau} \right) \right).
\end{align}

 
Here, $F_\mu$ denotes the electromagnetic force four-vector, $p_\mu$ is the particle momentum four-vector, $e,~m$ are the elementary charge and particle mass respectively, and $c$ is the speed of light. Equation \eqref{LAD} is derived for a point charge. Unphysical solutions appear, for example, when 
$F_\mu^{ext}=0$,  where in addition to the solution with a constant velocity, eq. \eqref{LAD} has a solution where the particle accelerates infinitely (the so-called "runaway solution"). The principle of causality is also violated here with pre-acceleration solutions - these solutions anticipate the change
of the force, so the particle accelerates before the force has been applied. The detailed explanation of these problems and suggestions for possible improvements are
given in \cite{rorlich_LAD,ArthurY}. 

However, even if these problems would be solved, the LAD equation is inconvenient for numerical integration. It is possible to integrate it backwards in time \cite{LAD_backwards}, 
but this is not applicable to many-body problems \cite{SokolovRenorm}. Therefore, the LAD equation, expressed by \eqref{LAD}, is inadequate for use in a simulation code.  

Numerous approximate models have been explored to eliminate the above-mentioned problems and to obtain a more efficient computational approach.
We consider a subset of these models appropriate for PIC implementation: B08 \cite{model1bell}, LL \cite{LLbook} (used also in \cite{model2zhidkov} and \cite{model2noguchi}), 
 S09 \cite{model3sokolov}, H08 \cite{Hededal}, and F93  \cite{model5ford}. We also consider the LL reduced (LLR) model \cite{TamburiniLL} where the spatio-temporal derivatives of the fields are discarded in the equation of motion because they are shown to have smaller contribution than the particle spin, that becomes important only in the quantum regime. Model B08 keeps only the leading-order term of the LL equation. The H08 model estimates the total energy radiated via the Larmour formula, and then this energy is discounted from the particle in the end of the timestep. All the models apply to the case of relativistic motion that is required to have an appreciable energy loss due to radiation emission. Their validity is limited to the classical domain, where the particle trajectory can still be considered to be a smooth function of time (the individual emission events do not take a great fraction of the particle energy, or in other words, the emitted energy in the Lorentz frame momentarily co-moving with the emitting particle is small compared with $mc^2$). Models LL and S09 also appear in Ref. \cite{qed_class_rr} that compares the asymptotic solution for equations of motion coming from strong field quantum electrodynamics with several classical equations of motion; their findings show that LL model is consistent with the asymptotic strong field QED description. Therefore, we will identify where the results obtained with other models are the same (or close enough) as with LL within the limits of the validity of the classical description.   
 
To facilitate the analysis for the PIC implementation, we will use the  3-vector form of the equations throughout. The total change of momentum in time depends on the Lorentz force ($\bold{F}_{L}$)   and the radiation reaction force ($\bold{F}_{RR}$):  
 \begin{equation}\label{all}
  \frac{d\bold{p}}{dt}=\bold{F}_L+\bold{F}_{RR} 
 \end{equation}
 where  $\bold{F}_L$, in CGS units, is given by:
 \begin{equation}\label{Lorentz}
 \bold{F}_L=e\left(\bold{E}+\frac{\bold{p}}{\gamma mc}\times\bold{B} \right).
 \end{equation}
The radiation reaction force ($\bold{F}_{RR}$) for all the considered models, in CGS units, is presented in Table 1. Here the radiation back reaction is explicitly given as an additional force acting on the particle, expressed as a function of the electromagnetic fields $\mathbf{E}$, $ \mathbf{B}$, the particle momentum $\mathbf{p}$, charge $e$, mass $m$ and relativistic factor $\gamma$, the speed of light $c$ and time $t$.

\begin{table*}[t!]
\begin{align}
\label{model1}& \text{\textbf{B08}}  & \left( \frac{d\bold{p}}{dt} \right)_{RR}  &  = -\frac{2}{3}\frac{e^4\gamma}{m^3c^5}\bold{p}\left(\bold{E_{\perp}}+\frac{\bold{p}}{\gamma mc}\times \bold{B}\right)^2\\[1em]
\label{model2} &\text{\textbf{LL\quad}}  & \left( \frac{d\bold{p}}{dt} \right)_{RR}  &  =
\frac{2e^3}{3mc^3} \Biggl\{ \gamma \Biggl( \biggl(\frac{\partial}{\partial t}+\frac{\bold{p}}{\gamma m}\cdot \nabla\biggr)\bold{E}+\frac{\bold{p}}{\gamma mc}\times\left(\frac{\partial}{\partial t}+\frac{\bold{p}}{\gamma m}\cdot \nabla\right)\bold{B}\Biggr) \Biggr. \nonumber \\ 
&&&\quad \biggl. +\frac{e}{mc}\Biggl(\bold{E}\times\bold{B}+\frac{1}{\gamma mc}\bold{B}\times\left(\bold{B}\times\bold{p}\right)+\frac{1}{\gamma mc}\bold{E}(\bold{p}\cdot\bold{E})\Biggr)
-\frac{e\gamma}{m^2c^2}\bold{p}\Biggl(\left(\bold{E}+\frac{\bold{p}}{\gamma mc}\times \bold{B}\right)^2-\frac{1}{\gamma^2m^2c^2}(\bold{E}\cdot\bold{p})^2\Biggr)\Biggr\}\\[1em]
\label{model3} &\text{\textbf{S09}} & \left( \frac{d\bold{p}}{dt} \right)_{RR} &=\frac{2e^3}{3m^2c^4}\frac{\bold{F_L}-\frac{1}{\gamma^2m^2c^2}\bold{p}(\bold{p}\cdot \bold{F_L})}{1+\frac{2e^2}{3\gamma m^3c^5}(\bold{p}\cdot\bold{F_L})}\times \bold{B}
-\frac{2\gamma e^2\bold{p}}{3m^3c^5}\left(\frac{\bold{F_L}-\frac{1}{\gamma^2m^2c^2}\bold{p}(\bold{p}\cdot \bold{F_L})}{1+\frac{2e^2}{3\gamma m^3c^5}(\bold{p}\cdot\bold{F_L})}\cdot \bold{F_L}\right)\\[1em]
\label{model4}&\text{\textbf{H08}}   &  \left( \frac{d\bold{p}}{dt} \right)_{RR}  & =-\frac{2}{3}\frac{e^4\gamma^3}{mc^3}\Biggl(\left(\bold{E}+\frac{\bold{p}}{\gamma mc}\times \bold{B}\right)^2-\frac{1}{\gamma^2m^2c^2}|\bold{p\cdot E}|^2\Biggr)\frac{\bold{p}}{p^2}\\[1em]
\label{model5}&\text{\textbf{F93}}  &  \left( \frac{d\bold{p}}{dt} \right)_{RR}  & =\frac{2}{3}\frac{e^2}{mc^3}\Biggl\{\gamma\frac{d\bold{F_L}}{dt}-\frac{\gamma}{m^2c^2}\frac{d\bold{p}}{dt}\times(\bold{p}\times \bold{F_L})
+\frac{1}{\gamma m^4c^4}\left(\bold{p}\cdot \frac{d\bold{p}}{dt}\right)\Bigl(\bold{p}\times(\bold{p}\times\bold{F_L})\Bigr)\Biggr\}\\[1em]
\label{model6}&\text{\textbf{LLR}}  & \left( \frac{d\bold{p}}{dt} \right)_{RR}  &  =\frac{2e^3}{3mc^3} \Biggl\{ 
 \frac{e}{mc}\Biggl(\bold{E}\times\bold{B}+\frac{1}{\gamma mc}\bold{B}\times\left(\bold{B}\times\bold{p}\right)+\frac{1}{\gamma mc}\bold{E}(\bold{p}\cdot\bold{E}) \biggl. \Biggr)
-\frac{e\gamma}{m^2c^2}\bold{p}\Biggl(\left(\bold{E}+\frac{\bold{p}}{\gamma mc}\times \bold{B}\right)^2-\frac{1}{\gamma^2m^2c^2}(\bold{E}\cdot\bold{p})^2\Biggr)\Biggr\}
\end{align}
\caption{Radiation reaction contribution to the equations of motion: $\mathbf{F_L}$ - Lorentz force; $\mathbf{p}$,  $e$, $m$ - particle momentum, charge and mass, $\gamma$ - relativistic factor; $\mathbf{E}, \mathbf{B}$ - electromagnetic fields, $c$ - speed of light, $t$ - time. }
\end{table*}

Particle-in-cell codes usually employ normalised units to bring all the quantities to similar orders of magnitude and express the physics as a function of fundamental plasma parameters. The normalisation in OSIRIS is as follows: $t\rightarrow t\omega_n$, $\mathbf{x}\rightarrow\mathbf{x}\omega_n/c$, $\mathbf{p} \rightarrow \mathbf{p}/mc=\gamma \mathbf{v}/c$, $\mathbf{E}\rightarrow e \mathbf{E}/mc\omega_n$, $\mathbf{B}\rightarrow e \mathbf{B}/mc\omega_n$. Here, $\mathbf{x}$ represents a vector in coordinate space, while $\omega_n$ is a chosen reference value for the normalising frequency, which, for instance, can be equal to the electron plasma frequency, or the frequency of the laser. In normalised units, the equations of motion of the particles are free of physical constants, except for a dimensionless coefficient 
\begin{equation}\label{damping_constant}
k=\frac{2\omega_n e^2}{3mc^3}
\end{equation}
which appears in the radiation reaction force term. Therefore, when including the radiation reaction force, the particle motion is no longer dependent only on the charge-to-mass ratio as in the Lorentz force $\mathbf{F}_L$. This poses an additional challenge for the PIC implementation of classical radiation reaction.

In standard PIC codes, the system is represented using macro-particles. Every macro-particle has the same charge-to-mass ratio as a single particle, and therefore the dynamics of the macro-particles is the same as the dynamics of the original particle species. However, after including the radiation reaction, this no longer holds. If we examine equations \eqref{all}, \eqref{Lorentz} and \eqref{model6}, we can see that the ratio between the radiation reaction force and the
Lorentz force is proportional to the cube of the charge, and to the reciprocal value of the squared mass
\begin{equation}
\frac{F_{RR}}{F_L} \propto \frac{e^3}{m^2}.
\end{equation} 

\begin{figure}[b!]
 \begin{center}
  \includegraphics[width=22em]{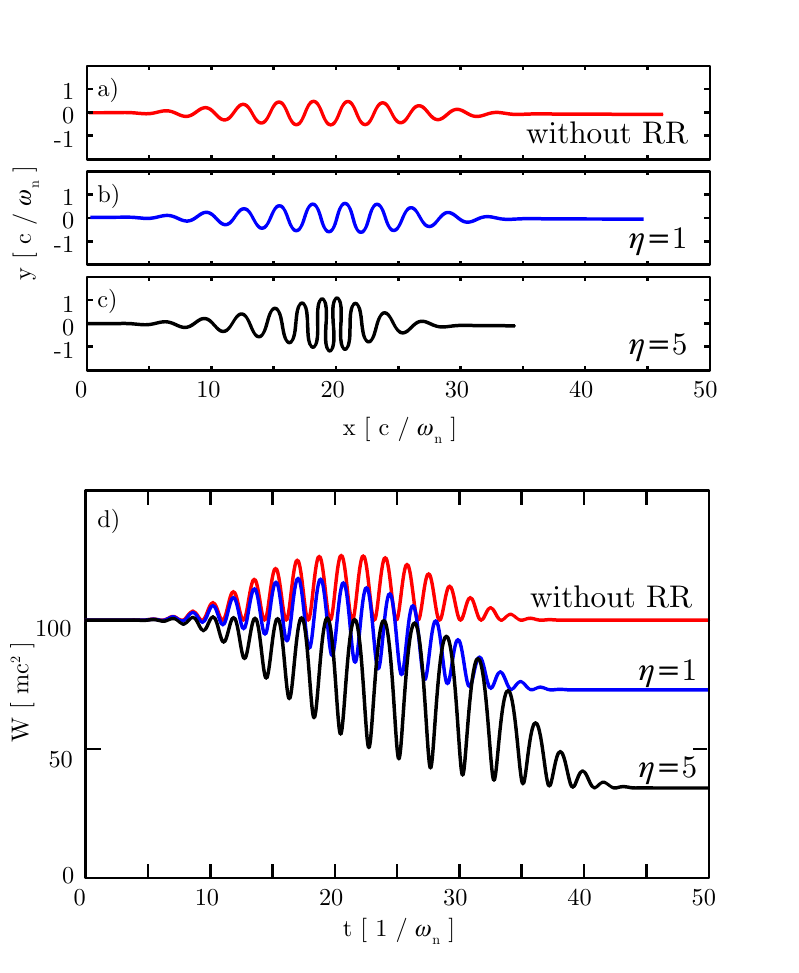}
  \end{center}
 \caption{ \label{macro} a) Trajectory of a macro-particle without radiation reaction b)-c) Trajectories of macro particles with weights $\eta=1$ and $\eta=5$ respectively, counter propagating with a laser pulse where $a_0=100$. Particle initial momentum is $p_0=-100$. d) Energy of the same macro particles from a)-c) as a function of time.}
 \end{figure}

Let us consider a macro particle that represents $\eta$ electrons. The charge of the macro particle is $e_m=\eta e$, and the mass is  $m_m=\eta m_e$.
For a single particle with the same mass and charge as the macro particle, the radiation reaction would be $\eta$ times stronger than in the case of a single electron:
 \begin{equation}
 \frac{F_{RR}}{F_L} \propto \frac{(\eta e)^3}{(\eta m_e)^2} = \eta \frac{e^3}{m_e^2}
 \end{equation}
and the trajectory of such particle would be different than the trajectory of a single electron (Fig. \ref{macro}). This result would be equivalent to assuming that $\eta$ electrons are radiating coherently. As a consequence, the results of a PIC simulation would be qualitatively different for different number of particles per cell or different cell sizes. To obtain the correct dynamics of a macro-particle, it is therefore essential to use the real charge and mass to calculate the correct radiation reaction coefficient for a particular particle species.  This approach yields the same result regardless of the macro-particle weight. 

\section{Comparison of the models with standard radiation mechanisms}

\begin{figure}
 \begin{center}
 \includegraphics[width=21em]{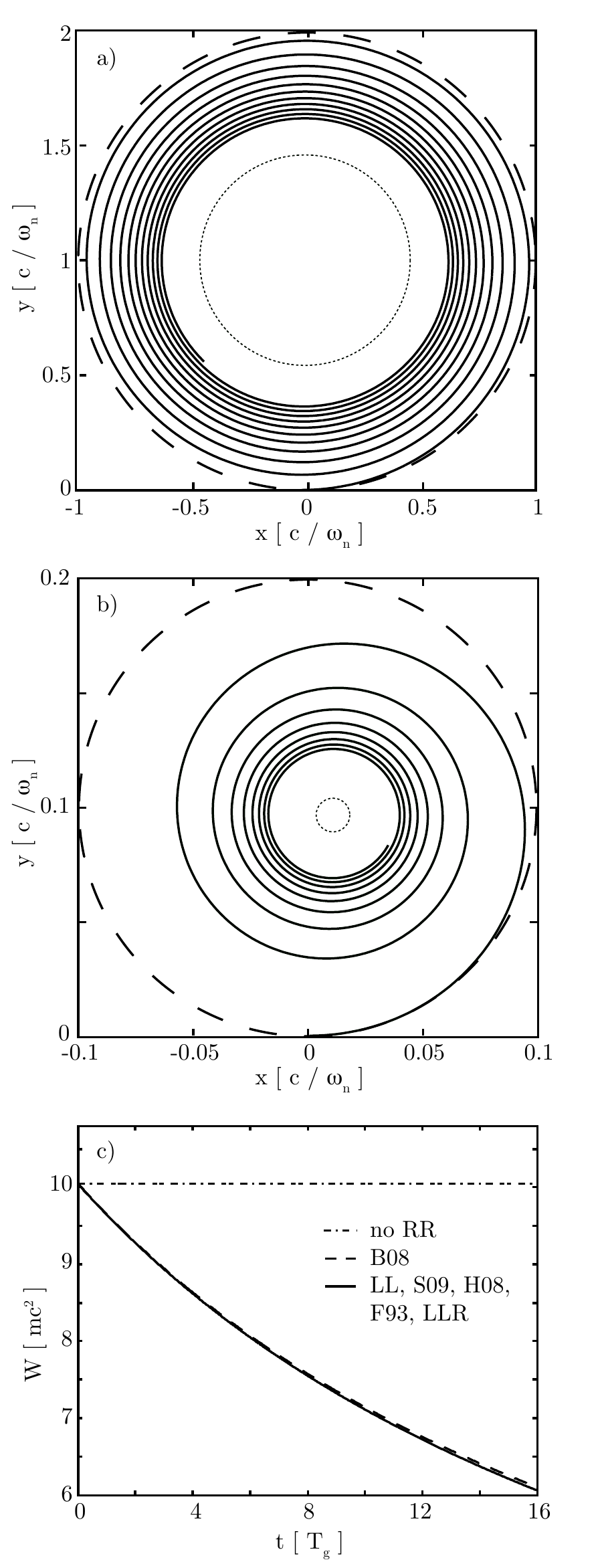}
 \end{center}
 \caption{\label{Synchro} Synchrotron radiation: electron trajectories and energy evolution. $\mathrm{T_g}$ - electron gyro period. The normalising frequency $\omega_n$ is the laser frequency corresponding to $\lambda = 1 \mathrm{\mu m}$. The initial conditions are: a) $p_0 = 100$, $B_0 = 100$, b) $p_0 = 100$, $B_0 = 1000$, c) $p_0 = 10$, $B_0 = 5975$. Dotted circle denotes the limit of the spiral motion. }
 \end{figure}
To examine the physics captured by the different models, we compare the dynamics of a single particle with well known examples for which the particle trajectory and radiated power are known. 

We first consider synchrotron radiation, where a particle
moves in a constant external magnetic field. Taking only the Lorentz force into account, we expect the trajectory to be a perfect 
circumference due to the $\bold{v} \times \bold{B}$ term. 
When the particle has a very high initial momentum, and is moving in an intense magnetic field, it radiates. The radiative energy loss over time is given by~\cite{LLbook}: 
\begin{equation}\label{en_loss_sync}
 -\frac{d\xi}{dt}=\frac{2e^4B^2}{3m^4c^7}(\xi^2-m^2c^4),
\end{equation}
where $\xi=\gamma mc^2$ is the total particle energy and $B$ is the magnetic field. If the particle is not relativistic, then $\xi \approx mc^2$ and the right-hand side of~\eqref{en_loss_sync} is close to zero, and therefore the energy loss is negligible. This is also the case when the
$\bold{B}$ field is small. However, an energetic particle in a high intensity magnetic field will lose a significant amount of energy over time. As the particle 
loses energy, its velocity decreases, and so does the curvature radius of its trajectory. This means that we do not expect a  
purely circular motion, but an inward spiral trajectory. When it loses a sufficient amount of energy, so that the right-hand side of~\eqref{en_loss_sync}
approaches zero \emph{i.e.} $\xi^2\simeq m^2c^4$, the particle trajectory then converges to a circumference. 
 
We have performed simulations for the same initial conditions using all the considered models (eqs. \eqref{model1}-\eqref{model6}). A typical example for a very strong damping is shown in Fig. \ref{Synchro} where we can see that all the energy loss predictions closely resemble each other and match the analytical expression \eqref{en_loss_sync}. The strongest difference to eq. \eqref{en_loss_sync} can be seen for the model B08 \cite{model1bell}, which is still smaller than $0.3 \%$. Therefore, 
we can confidently state that this effect is well resolved by all the models. This is not surprising because in this configuration, where $\bold{B}$ is constant and $\bold{B} \perp \bold{p}$, equations \eqref{model1}-\eqref{model6} reduce to:

\begin{equation}\label{reducedsynchro}
\begin{tabular}{ll}
 \textbf{B08} &$\left( \frac{d\bold{p}}{dt} \right)_{RR}=-kB^2\frac{p^2}{\gamma}\bold{p}$\\[1em]
 \textbf{LL, S09, LLR,   } &  $ \left( \frac{d\bold{p}}{dt} \right)_{RR} =-kB^2\frac{p^2+1}{\gamma}\bold{p}.$ \\
 \textbf{H08, F93   } & \\[1em]
\end{tabular}
\end{equation}
For a highly relativistic particle, $p\gg1$ and $p^2+1\approx p^2$, so these two equations \eqref{reducedsynchro} are expected to yield similar results.

\begin{figure}
\begin{center}
 \includegraphics[width=20em]{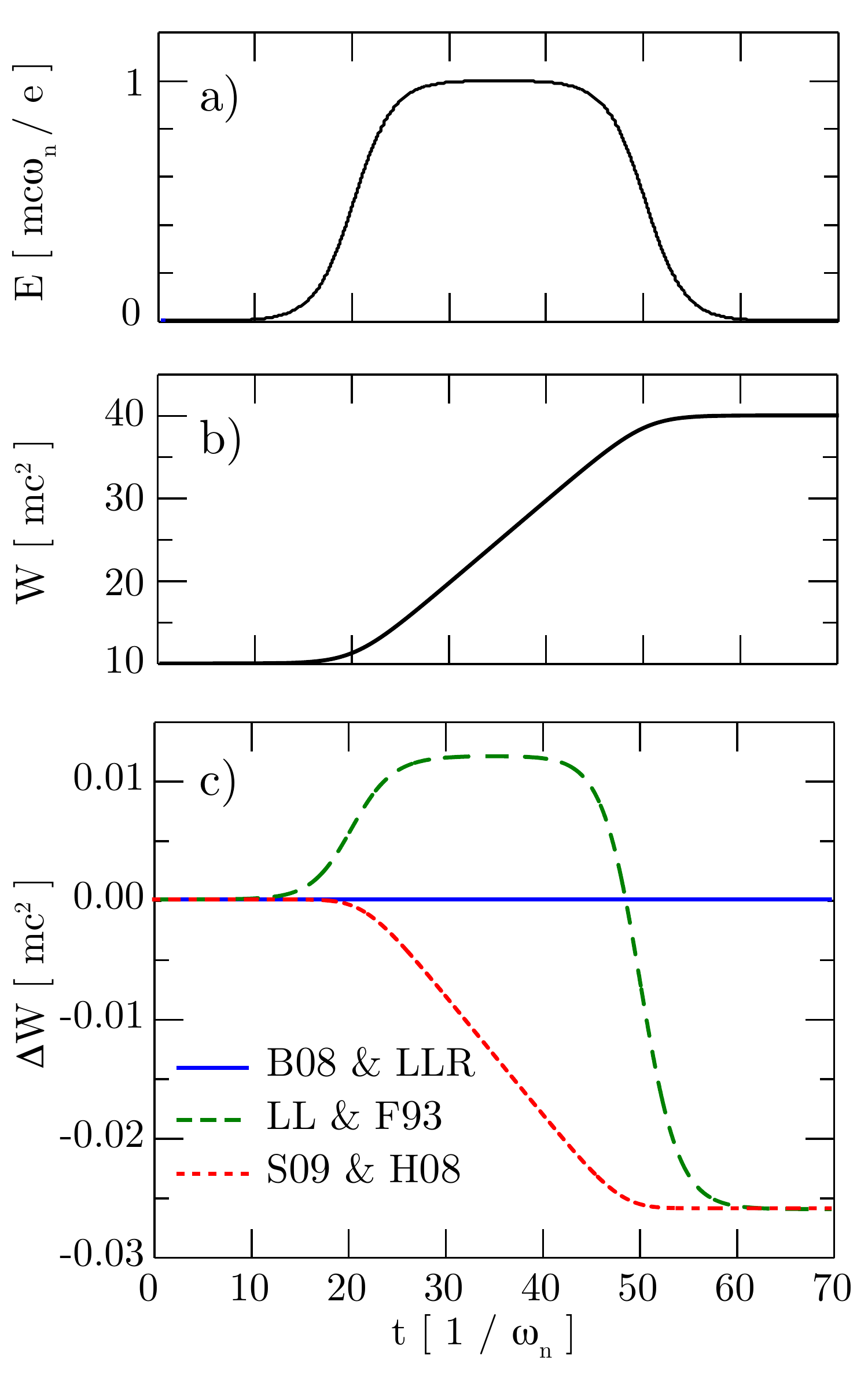}
\end{center}
 \caption{\label{Bremss} a) Electric field as felt by the particle over time; b) total particle energy vs. time; c) the difference in energy compared to the case without radiation reaction for k=0.001.}
 \end{figure}
 
Another illustration of the role of radiation reaction is Bremsstrahlung radiation. We let a charged particle propagate ($\gamma_0=10$) along the x-axis starting from the origin in a static external electric field parallel to the particle motion given by (in normalised units):  
\begin{equation}\label{Eprofile}
E(x)=\frac{E_0}{2}\left(\frac{e^{x/a}-e^{-x/a}}{e^{x/a}+e^{-x/a}}+\frac{e^{(L-x)/a}-e^{-(L-x)/a}}{e^{(L-x)/a}+e^{-(L-x)/a}}\right)
\end{equation}
where $E_0=1$, $a=4$ and $L=30$ (see Fig. \ref{Bremss}).
The dominant term in eqs. \eqref{model1}-\eqref{model6} that leads to essentially the same result for the synchrotron radiation \eqref{reducedsynchro}, is identically equal to zero in this specific configuration. This allows us to explore the conceptual differences between some of the models. 
In Fig. \ref{Bremss} we observe that the LL model and the F93 model depend on the gradient of the applied electric field. There is no radiation reaction observed in the B08 and the LLR models. The S09
 and H08 models show some energy loss, but the rate at which the particle loses energy is 
 constant where the electric field is constant and is not affected by the field gradient.

This is better understood if we reduce the equations in Table 1  to the case where $\bold{E} || \bold{p}$: 
\begin{equation}\label{reduced}
\begin{tabular}{ll}
 \textbf{B08} and \textbf{LLR} &$\left( \frac{d\bold{p}}{dt} \right)_{RR}=0$\\[1em]
 \textbf{LL} and \textbf{F93  } & $ \left( \frac{d\bold{p}}{dt} \right)_{RR} =k\gamma(\frac{\partial\bold{E}}{\partial t}+\frac{\bold{p}}{\gamma}\frac{\partial\bold{E}}{\partial x})$ \\[1em]
 \textbf{S09} & $\left( \frac{d\bold{p}}{dt} \right)_{RR}=-k\frac{\bold{E}^2\bold{p}}{\gamma+k\bold{p}\bold{E}}$\\[1em]
 \textbf{H08} & $\left( \frac{d\bold{p}}{dt} \right)_{RR}=-k\gamma \frac{\bold{E}^2\bold{p}}{\bold{p}^2}$.
\end{tabular}
\end{equation}

We see that the LL and F93 models reduce to the same expression. Their radiation reaction force depends only on the field derivatives, not on the absolute value of the field. There is no effect associated with the radiation reaction force in the B08 or in the LLR models. The S09 and H08
  models lead to expressions that do not depend on the field derivatives, but only on the absolute value of the
  field, as can be observed in Fig. \ref{Bremss}.
 
This analysis shows that if we choose one of the models that are insensitive to the rate of the field 
change, we can apply it exclusively in a regime where this change is small enough not to significantly influence the motion.
The condition that must be satisfied is then:
\begin{equation}\label{validity_brems}
\frac{\bold{p}}{m^2c^2}~| \bold{F}_\perp |^2\gg\frac{d\bold{F}}{dt}, 
\end{equation}
where $\mathbf{p}$ is the particle momentum, $\mathbf{F_\perp}$ is the perpendicular component of the Lorentz force $\mathbf{F}_L$ with respect to $\mathbf{p}$.

To show that the contribution of the longitudinal fields to the radiation reaction in different models will be small in the vast majority of scenarios, 
we have chosen to illustrate a case with a very strong field $E_0=0.2~ E_{C}$ (Fig. \ref{Bremss}c) in a setup similar as in \cite{model5ford}. Here $E_{C}$ stands for the Schwinger limit  ($E_C=m^2c^3/e \hslash $), which marks the transition from the classical to the quantum regime. 
For a field amplitude of $E_0=0.2~ E_{C}$ the difference in the energy loss predicted by different models is smaller than 0.1 \% compared with the total electron energy (Fig. \ref{Bremss}b).
Since at $E\simeq E_C$ the classical models cannot be applied, a field amplitude outside the scope of applicability of the models is necessary to observe a non-negligible difference between the models. 

Additionally, we would like to stress that the applicability condition \eqref{validity_brems} is satisfied in most physical scenarios of interest. In particular, for laser pulses, the inequality \eqref{validity_brems} is satisfied if $a_0\gamma\gg 1$, which is true whenever radiation reaction is significant. In \cite{TamburiniLL} this was confirmed by 
comparing the contribution of the particle spin and the contribution of radiation reaction force arising from $d\mathbf{B}/dt$ and $d\mathbf{E}/dt$ in the plane wave scenario. In this comparison, the spin gives a bigger 
contribution. In the classical regime that we are addressing here, the spin contribution is negligible, and so are the contributions of  $d\mathbf{B}/dt$ and $d\mathbf{E}/dt$. 

In summary, any of the proposed models can be used to describe the classical radiation reaction dominated regime.

\section{Testing the role of classical radiation reaction with the dynamics of electrons in intense laser pulses}

In order to examine the role of classical radiation reaction in scenarios with intense laser pulses and to determine the conditions where such models should be used we consider the dynamics of a single electron interacting with a laser pulse. This is one of the main scenarios where radiation reaction can be explored \cite{QED1,QED2} and of very high relevance for future laser facilities \cite{model1bell, model3sokolov, TamburiniLL, First_PRL,Ruhl_threshold, capturePiazza, esarey_relmass}. 
The laser pulse normalised vector potential is written as \cite{Gibbon}:   
\begin{equation}\label{slow_vary}
\bold{A}(x,t)=a_0 f(\phi) \cos \phi \ \mathbf{e}_y
\end{equation}
where the temporal envelope $f(\phi)$ is a slowly varying function relative to the laser cycle ($df/dt\ll\omega_0 f$), and the phase of the wave is given by $\phi = \omega_0 t-\omega_0 x/c$. We choose for $f(\phi)$ a polynomial 
function $10\tau^3-15\tau^4+6\tau^5$ where $\tau=\phi / (\omega_0~ t_{\mathrm{FWHM}})$ takes values in the domain [0,1] to define the envelope rise for $0<\phi /\omega_0<t_{FWHM}$. Here, the pulse duration $\tau_{\mathrm{FWHM}}$ is defined as full-width-at-half-maximum in the laser fields, and the envelope profile is symmetric relative to the point of the peak intensity located at $\phi=\omega_0~t_{FWHM}$. The relativistically invariant normalised vector potential $a_0$ can be related to the linearly polarised laser intensity through $a_0=0.8 \sqrt{I[10^{18}~\mathrm{W/cm^2}]} \lambda [\mu \mathrm{m}]$.

\begin{figure}[b!]
 \begin{center}
  \includegraphics[width=22em]{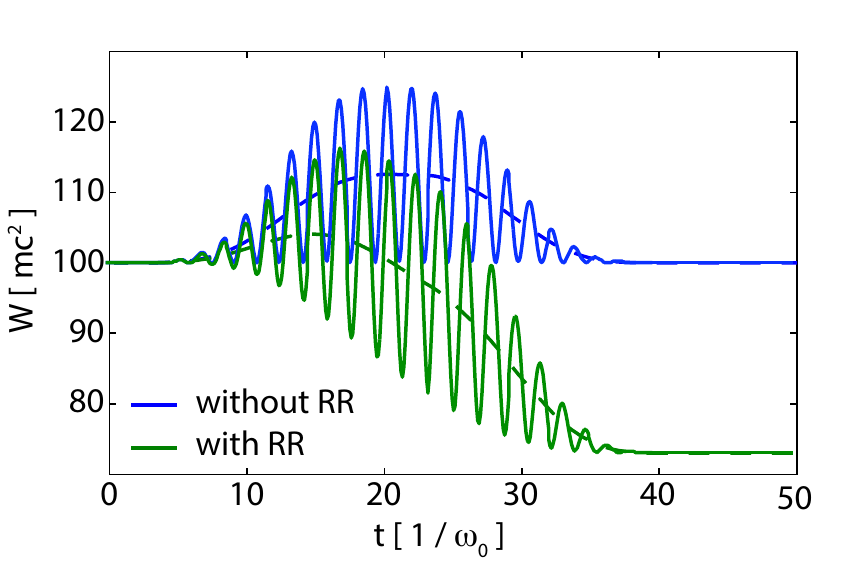}
  \end{center}
 \caption{ \label{CompareLC} Particle energy over time in the field of linearly (solid line) and circularly (dashed line) polarized laser pulse.}
 \end{figure}

In the field of a linearly polarized laser pulse, a charged particle undergoes quiver motion. Without the radiation reaction force, the total particle energy remains unchanged
after the interaction with the laser pulse. With radiation reaction, the total energy can decrease substantially. If we consider a circularly polarized laser pulse, the situation is similar. In Fig. \ref{CompareLC} this is illustrated for a linearly and a circularly polarized laser with identical temporal envelopes and the same intensities ($a_0^{LP}=100$, $a_0^{CP}=100/\sqrt{2}$, $\lambda_0=1\rm{\mu m}$). The initial  normalised momentum of the particle is $p_0=100$, opposite to the laser propagation direction.  The total energy that the particle loses while interacting with the laser is about
30\% and is the same for the linearly and the circularly polarized case. All the models give similar results (the biggest difference in the final energy in Fig. \ref{CompareLC} is 0.03\%).  

It is important to estimate the threshold when radiation reaction starts to significantly affect the motion. In Fig. \ref{CompareLC}, we see that for counter-propagating particle and a laser pulse there is already strong radiation reaction for $a_0=100$ if $p_0=100$. However, if we repeat the simulation 
with the same laser parameters for a less energetic particle, radiation reaction has a weaker effect. For a particle starting at rest,
there is no radiation reaction observed  for $a_0=100$. This indicates that in order to estimate whether radiation reaction is important for a set of particular conditions, it is  
also required to have the information about the relative motion of the particle and the laser. We can assume that the
radiation reaction force is significant when the following condition is satisfied: 
\begin{equation}\label{threshold1}
\frac{|\mathbf{F}_{rad}|}{|\mathbf{F}_L|}>10^{-3}.
\end{equation}
In eq. \eqref{model2}, the dominant term due to radiation for high $\gamma$ (ultra-relativisic particle) written in units where all quantities are normalised to the laser frequency is: 

\begin{equation}
-\frac{2}{3}\frac{e^2\omega_0}{mc^3}\gamma\mathbf{p}\left( \left( \mathbf{E}+\frac{\mathbf{p}}{\gamma}\times \mathbf{B}\right)^2-\frac{1}{\gamma^2}\left( \mathbf{E}\cdot \mathbf{p}\right)^2 \right).
\end{equation} 

In these units $E, B\approx a_0$ and
$\gamma \approx p$, and thus the left hand side of eq. \eqref{threshold1} is on the order of
$
(2e^2\omega_0/(3mc^3))\gamma^2 a_0
$
for a particle counter-propagating with the wave.
For a laser with a wavelength $\lambda_0\approx 1\mathrm{\mu m}$ eq. \eqref{threshold1} then becomes:
 \begin{equation}\label{threshold3}
 \left(\frac{1\mathrm{\mu m}}{\lambda_0} \right)\left(\frac{a_0}{10} \right)\left(\frac{\gamma_0}{10^2} \right)^2\gtrsim 1.
 \end{equation}

 The limits of validity of the classical approach for radiation reaction have been discussed
 in \cite{Ruhl_threshold, SokolovLimits}. When the electric field of the laser approaches
 the Schwinger field in the rest frame of the particle, quantum effects are expected
 to dominate. For a head-on collision, where radiation reaction has the strongest effect on the motion, 
 the classical equations can then be used if 
  \begin{equation}\label{threshold4}
 \left(\frac{1\mathrm{\mu m}}{\lambda_0} \right)\left(\frac{a_0}{10} \right)\left(\frac{\gamma_0}{10^4} \right)\ll 1.
 \end{equation}
Therefore, from eq. \eqref{threshold3} and eq. \eqref{threshold4} we can identify the range of applicability of the classical radiation reaction models
described here for the case of an electron colliding head-on with a laser as:
  \begin{equation}\label{threshold5}
 \frac{10^4}{\gamma_0^2} \lesssim \left(\frac{1\mathrm{\mu m}}{\lambda_0} \right)\left(\frac{a_0}{10} \right)\ll \frac{10^4}{\gamma_0}
 \end{equation} 
For instance, a 300 J laser pulse with $\lambda=1\mu \rm{m}$ and 30 fs duration, yields a peak power of 10 PW and, when focused to a $\mathrm{10\ \mu m}$ focal spot, a peak intensity of $I =10^{22}~\mathrm{W/cm^2}$. The corresponding vector potential is $a_0\approx 100$.
Future laser facilities will be able to provide normalised vector potentials of this magnitude, and for $\gamma_0\gtrsim100$ the inequalities \eqref{threshold5} are verified.


\section{The role of the timestep for simulating particle motion in an intense electromagnetic wave}

The dynamics of electrons in intense electromagnetic waves can be strongly relativistic. At ultra high intensities the relative dynamics of the particles in the laser field is highly nonlinear and the relevant numerical parameters must be selected with extra care. In this section, we will see that the choice of the simulation timestep is very important for the accuracy of the final result, and that the optimal value depends on the laser intensity, electron energy, total simulated time, as well as whether radiation reaction is significant or not. 

The higher resolution requirements at $a_0\gg1$ were reported previously in \cite{Alexey_subcycling}, where they studied the particle motion in an intense electromagnetic plane wave in 1D geometry and found the approximate convergence condition $c \Delta t/\lambda\ll 1/a_0$ (here $\lambda$ and $a_0$ are the laser wavelength and the normalised vector potential, $c$ is the speed of light and $\Delta t$ the simulation time step). Their study focuses on long interaction at relativistic intensities with $a_0\sim10$. The higher resolution requirement is even more relevant in scenarios where radiation reaction plays an important role because of the prospects of dealing with ultra-high intensities ($a_0\gg10$). However, the state-of-the-art and near-future laser facilities that aim at reaching such intensities are usually short ($\sim$ 30 fs), so we opt to consider short laser duration in our further analysis. In fact, as discussed below, we will have different resolution requirements depending on the particle energy, momentum direction and the maximum intensity of the laser field. For our case, the resolution requirements could be less strict than that of ref. \cite{Alexey_subcycling} depending on the energy of the interacting particles. As the average energy loss due to radiation reaction also depends on the laser duration \cite{First_PRL}, we could expect different resolution requirements for different laser durations in cases where there is significant radiation reaction.  

An initially small integration error can, in principle, grow during a long simulation. To assess this effect, we perform temporal resolution tests for realistic simulation timescales using the fourth order Runge-Kutta integration method. Usually, the longest simulations that can be performed using PIC codes are the ones that can take advantage of the moving window technique used to simulate only a portion of plasma around the laser pulse \cite{Moving_window}. We therefore take a typical laser wakefield acceleration (LWFA) setup for a 0.5 GeV electron acceleration stage operating in matched conditions \cite{LWFAscailing} as an example of one such long simulation. For a driver laser wavelength $\lambda_{LD} =1~ \mu \mathrm{m}$, it is required to have a 1.6 mm plasma length, laser $a_{0D} \simeq 1.6$ and the ratio between the laser and the electron plasma frequency  $\omega_{0}/\omega_{pe}\simeq 10$. This amounts to a total simulation time of $T_\mathrm{total}\simeq10^4~\omega_{0}^{-1}$, where the laser frequency is given by $\omega_{0}=1.88\times10^{15}~\mathrm{s}^{-1}$. We take $T_\mathrm{total}$ to be a characteristic time of a long simulation. The conclusions drawn by using $T_\mathrm{total}$ will be applicable for any simulations where simulation time $T_\mathrm{sim}\lesssim T_\mathrm{total}$, which applies to most PIC simulations performed with currently available resources.

Now that we established a typical simulation time, we consider the motion of an electron in a transversely plane electromagnetic wave with a longitudinal envelope that propagates along the x-direction.  The electromagnetic wave is expressed analytically according to the eq. \eqref{slow_vary} and has $30\  \mathrm{fs}$ duration at FWHM in intensity. The electron is considered as a test-particle that gives no feedback to the fields, and its motion is followed always for the same amount of time $T_\mathrm{total}$.  The electron initial momentum is chosen to be parallel to the wave propagation direction and either co- ($p_0>0$) or counter-propagating ($p_0<0$) with the laser.  We have studied the convergence of the particle trajectories for different initial momenta $p_0$, and different vector potential of the wave $a_0$ varying the timestep, while keeping the simulation duration in the units normalised to the laser frequency $\omega_0$ equal to $T_\mathrm{total}$. 

For some examples when $p_0>0$ the total interaction time $T_\mathrm{int}$ of the particle with a laser may be much longer than the time we are interested in $T_\mathrm{total}$. Hence, for our purpose the resolution required for a reliable simulation ensures correct numerical integration of trajectories of such particles within the simulation time $T_\mathrm{total}$. We would like to warn the reader that this resolution may be insufficient if one wishes to simulate the entire duration of laser-electron interaction if $T_\mathrm{int}\gg T_\mathrm{total}$ .

\begin{figure*}[t!]
\begin{center}
 \includegraphics[width=44em]{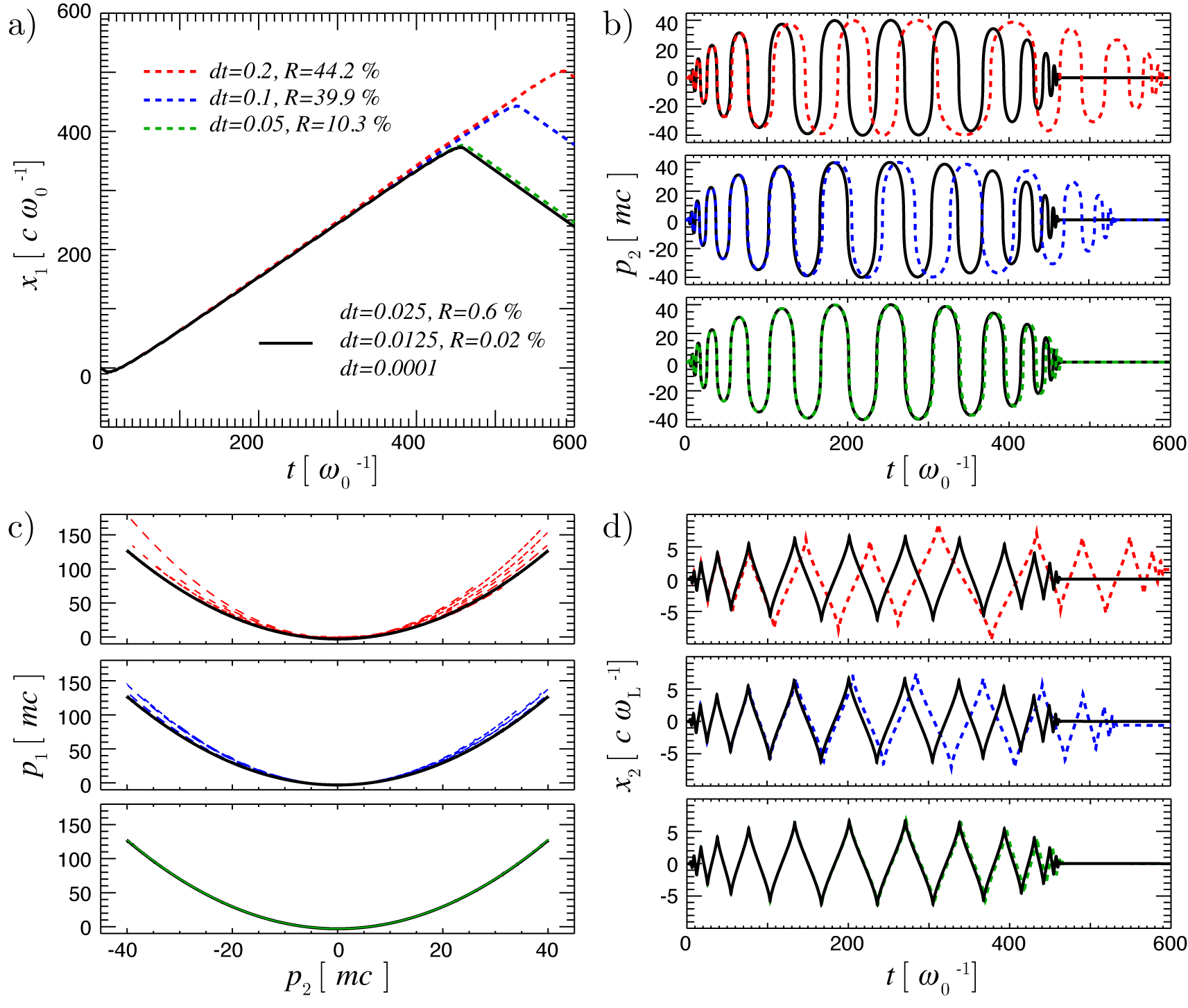}
 \end{center}
 \caption{\label{trajectories} Illustration of converged and unconverted particle trajectories. a) Longitudinal coordinate over time, b) Transverse momentum over time, c) Longitudinal  as a function of transverse momentum, d) Transverse coordinate over time. c) - d) Upper plots show the difference of the particle trajectory obtained with $dt=0.2$  (dashed red line) compared with the converged case (solid black line). Middle plots show the same comparison for $dt=0.1$ (blue dashed line), and lower plots for $dt=0.05$ (green dashed line).  }
 \end{figure*}

In order to establish a quantitative convergence criteria that would allow an automatic evaluation, we have first performed the simulations with a very small time step $dt=10^{-4}~ \omega_0^{-1}$, well beyond the convergence limit for each individual case. Then, we chose 1000 reference points equally spaced in time (for T=10, 20, ...) along the trajectory which would serve as a benchmark to compare with the runs with longer $dt$. We can then define the relative error as: 
\begin{equation}
R=\frac{\sum |\Delta \gamma|}{\sum \gamma}.
\end{equation}
Here, $\gamma$ is the Lorentz factor of the converged result, $|\Delta \gamma|$ stands for the absolute difference between this value and the one from the current test-particle run and the sums are over all reference points. We take the error to be acceptable if $R <$1\% for which we consider that the result has converged. An illustration of converged and unconverted particle trajectories for $a_0=40$ and $p_0=-3$ is shown in Fig. \ref{trajectories}. The figure shows  that the condition $R <$1\% is conservative because even a small visible deviation in a particle trajectory leads to $R \gg$1\%. 

\begin{figure*}[t!]
\begin{center}
 \includegraphics[width=42em]{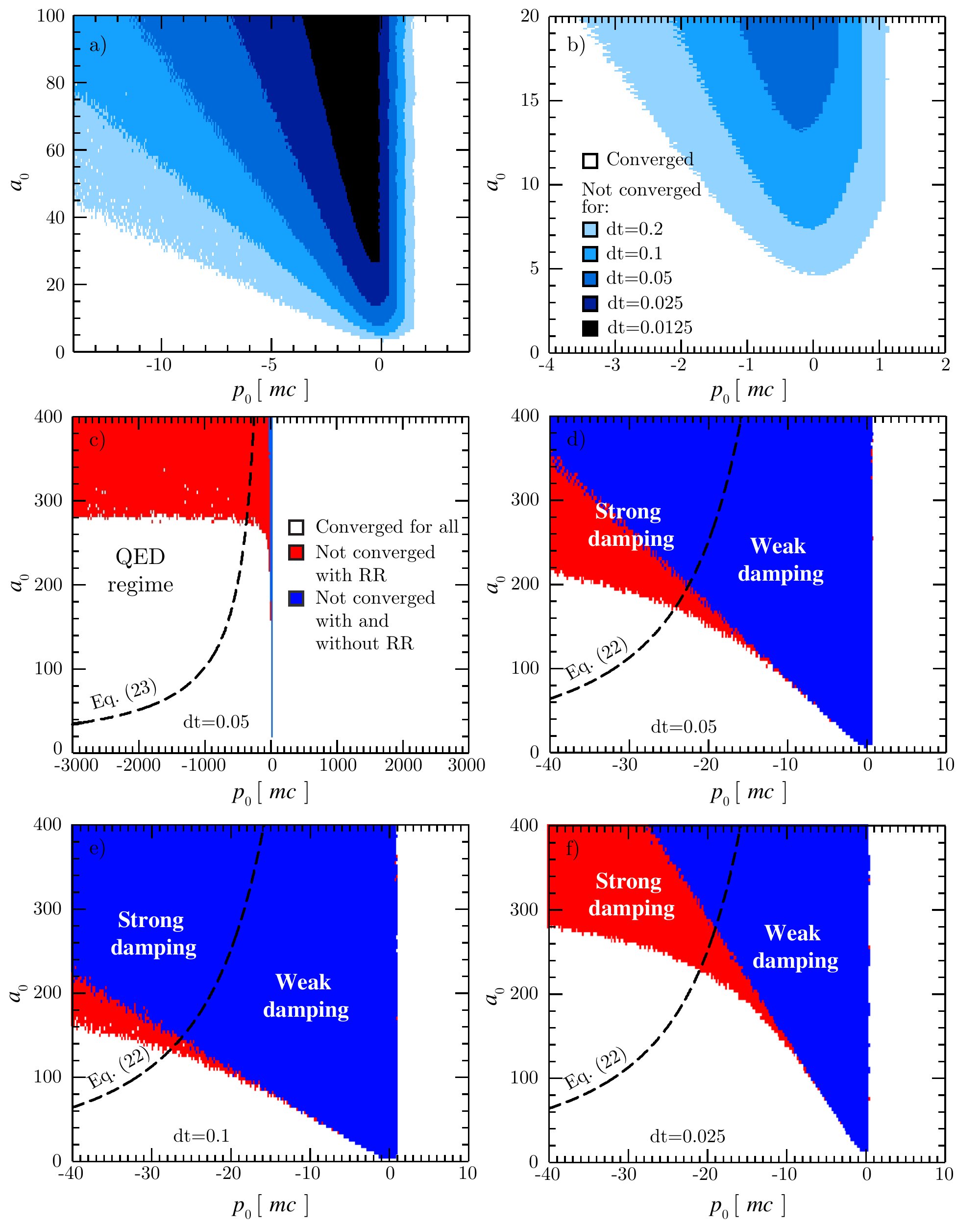}
 \end{center}
 \caption{\label{timestepconv} Convergence map for different time steps at a) high and b) low intensities.  Convergence map for c), d) $dt=0.05$, e) $dt=0.1$, f) $dt=0.025$, with and without accounting for radiation reaction. Color legend for both a) and b) is presented in panel b), while for c) - f) it is presented in panel c). The laser pulse duration was 30 fs ($\simeq 56.5~ \omega_0^{-1}$), and total simulation time for all simulations $T_\mathrm{total}=10^4~ \omega_0^{-1}$. Below $a_0=100$, the convergence does not depend on whether radiation reaction is included or not. Below $a_0=5$, the standard condition (dt=0.2 $\sim$ 30 points per laser period) is enough. For higher values, the most limiting case is laser in a cold plasma - the interaction of ultra relativistic particles can still be modelled with $dt=0.2$.  The difference in the regions with higher values of $a_0$ with and without radiation reaction can be attributed to the particle energy loss that lowers the effective $p$ of the particle.}
 \end{figure*}

\begin{figure*}[t!]
\begin{center}
 \includegraphics[width=44em]{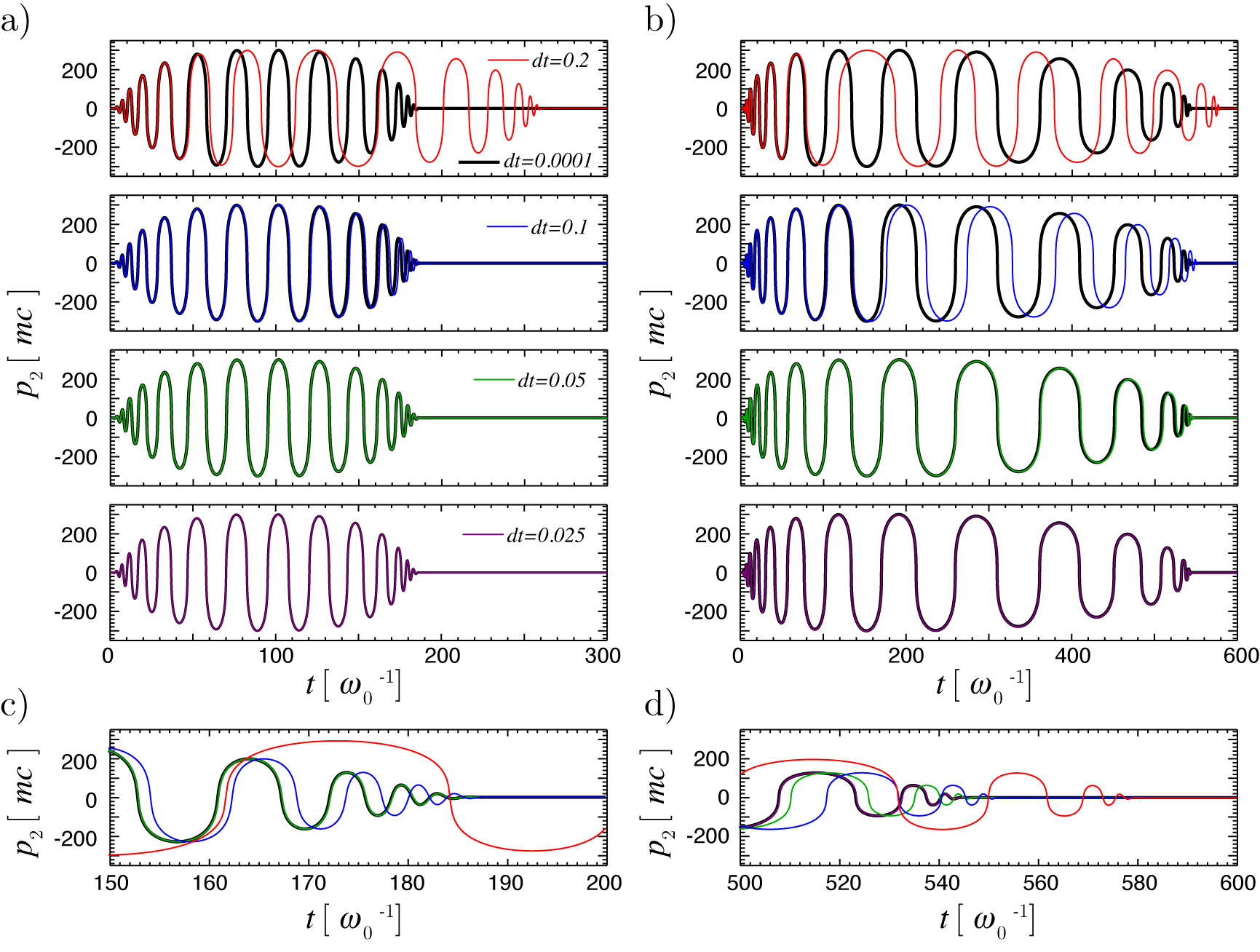}
 \end{center}
 \caption{\label{trajectories_rrconv} Particle trajectories for $a_0=300$ and $p_0=-40$. Transverse momentum $p_2$ versus time a) without and b) with radiation reaction. Thick black lines correspond to the fully converged trajectories.  The green lines corresponds to $dt=0.05$, for which the simulation has converged without radiation reaction and has not converged when radiation reaction is taken into account. This is even more evident when we compare all the trajectories on a smaller time interval near the end of the interaction c) without and d) with radiation reaction. }
 \end{figure*}

Comprehensive tests were done for a wide range of values of $a_0$ and initial momenta. A subset of these is plotted in Fig.  \ref{timestepconv}, where each pixel represents a single simulation (300$\times$100 pixels in each panel). It can be seen that below $a_0=5$, the condition of having $\sim$ 30 points per laser period is sufficient (here equivalent to $dt$=0.2 $\omega_0^{-1}$). If  $a_0$ is higher, the most demanding case is when the particle has very low energy before the interaction with the laser. In that case, the particle can get a strong kick, on the order of its total energy, within a single time step. This can modify its direction of motion and put the particle on a different trajectory, allowing the error to grow as the particle gains energy in the laser field. If a particle is already ultra-relativistic, the same error in the accelerating force does not change the motion significantly. 

Our findings are consistent with the ones reported in Ref. \cite{Alexey_subcycling}. They have shown that the "stopping point" of electron trajectories is the most sensitive section with respect to integration errors. Resolution requirements are therefore expected to be more strict for particles that do have a point of zero-momentum during the laser interaction. 

From Fig. \ref{timestepconv}a we can conclude that for studying the interaction of relativistic electron beams with intense lasers where $\gamma>a_0/3$, a resolution of 30 timesteps per laser period can still be kept. For smaller values of initial $\gamma$ this no longer holds and  special care should be applied when simulating intense lasers interacting with cold plasmas. These are general conclusions that apply whether or not the radiation reaction is accounted for in the code. We have repeated the same procedure with and without radiation reaction, and for lasers where $a_0<80$ we have not observed any differences in the convergence map (Fig. \ref{timestepconv} a,b). 

A difference between the convergence map with and without RR (Fig. \ref{timestepconv} c-f) arises only for very high $a_0$. This is not surprising since in these cases the particles lose significant energy due to the strong radiation reaction, and thus move to the regions in the parameter space of Fig. \ref{timestepconv} where smaller $dt$ would be required. For instance, if at the start of the interaction with a laser of $a_0=300$, an electron has $p_0=-40 ~mc$, after the interaction it will have $p=-15~ mc$. Figure  \ref{timestepconv}e shows that  $dt=0.1$ is not small enough for the particle trajectory to converge, with or without radiation reaction (the corresponding trajectories are illustrated in Fig. \ref{trajectories_rrconv} coloured blue). The electron trajectory satisfies our convergence criteria without radiation reaction for $dt=0.05$, but not if we take radiation reaction into account (Fig. \ref{timestepconv}d and Fig. \ref{trajectories_rrconv} green). However, for $p_0=-15$, the convergence criteria without radiation reaction is also not satisfied for dt=0.05 because lower particle momentum requires higher resolution (it is easier to reach the "stopping point" with a lower momentum). This means that by causing energy loss, and thus lowering the particle momentum, radiation reaction leads to more stringent convergence criteria that is evident in Fig. \ref{timestepconv} c-f. If we consider the lowest value of the average momentum in the simulation (i.e. the final momentum), we can apply the convergence map obtained without radiation reaction. A way to quickly estimate the total energy loss during a head-on collision with a laser pulse can be found in Ref.  \cite{First_PRL}.

\section{Computational overhead of radiation reaction modelling}

\begin{figure}
\begin{center}
 \includegraphics[width=22em]{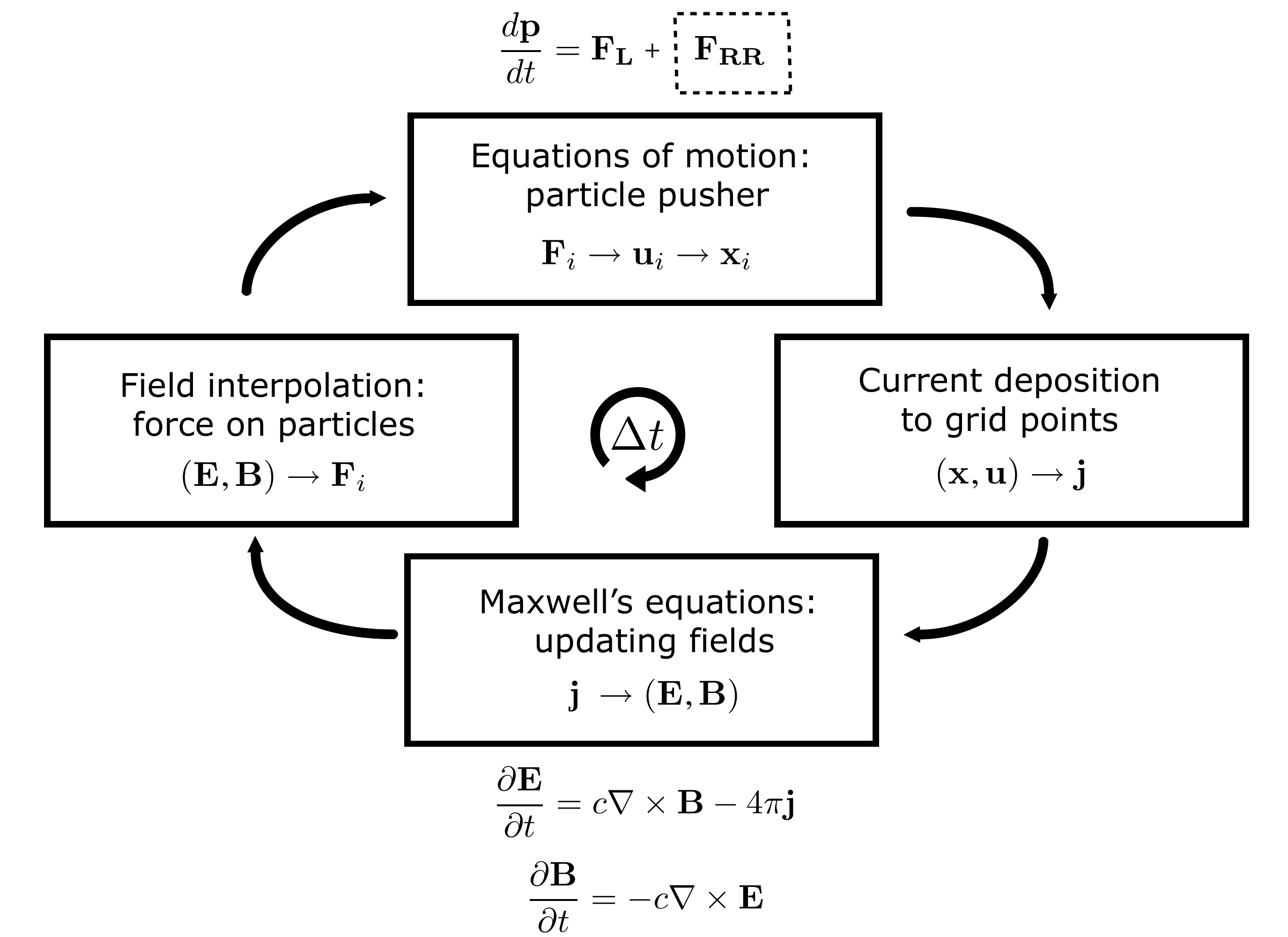}
 \end{center}
 \caption{\label{PICloop} Main PIC loop. $\mathbf{x}_i$, $\mathbf{u}_i$, $\mathbf{F}_i$ - particle position, generalized velocity $\mathbf{u}_i=\gamma \mathbf{v}_i$ and force acting on the $i$-th particle; $\mathbf{j}$-total current; $\mathbf{E},\ \mathbf{B}$ the electric and the magnetic field respectively.}
 \end{figure}

To assess the computational overhead of the radiation reaction schemes for particle-in-cell codes, we need to consider the full structure of the PIC loop (presented in Fig.  \ref{PICloop}) and the main aspects that affect its performance.  The electromagnetic fields in the simulation are stored on the grid nodes, while particles explore the full 6D phasespace. The fields stored on the grid are interpolated to the particle positions and then the particles are pushed according to the Lorentz force. Later, the currents created by all the particles are deposited on the grid, where this information allows to advance the fields on the grid nodes. The new fields are then interpolated again to the new particle positions to start the next iteration. 
OSIRIS is parallelised through a spatial domain decomposition, where only the neighbouring computer nodes are required to exchange information. This exchange consist of sending the information about the particles that moved to the sector of space that belongs to the neighbouring node and about the electromagnetic fields in the vicinity of the boundary between the two. 

Introducing the radiation reaction into the PIC loop, in principle, does not affect the parallelisation and should not affect the overall scalability of the PIC code. The main difference is in the particle pusher, where instead of the Lorenz force, it is necessary to apply a full equation of motion that also incorporates the radiation reaction force. The particle pusher in OSIRIS can follow a second-order Boris method \cite{book_simulation}, or a fourth-order Runge-Kutta integration method for the particle equations of motion. Radiation reaction can be implemented in both of them (see ref. \cite{TamburiniLL} for instructions on adapting the LL method for Boris pusher).

When examining the equations \eqref{model1}-\eqref{model6} one can expect a significant overhead of the particle
pusher as compared to the case where we only had the Lorentz force. 
To estimate the additional computational cost, we have determined the particle pusher overhead  for every model used here.  
The overhead is defined as the ratio between the number of floating point operations (flop) required for the radiation 
reaction force and for the Lorentz force respectively.
\begin{table}
\begin{align}
&\text{Model}& \text{Overhe}&\text{ad}&  \text{Added quantities}& \nonumber\\
&\text{B08}&1&&\text{0}& \nonumber\\
&\text{LL}& 4.5&& \text{6 per grid point}&\nonumber\\
&\text{S09}& 2&&\text{0}& \nonumber\\
&\text{H08}& 1.5&&\text{3 per particle}& \nonumber\\
&\text{F93}& 2&&\text{3 per particle}& \nonumber\\
&\text{LLR}& 2&&\text{0}& \nonumber
\end{align}
\caption{Computational cost for each RR model}
\end{table}

In state-of-the-art supercomputers, memory access is often more time-consuming than 
additional flop. The balance depends on a particular computer architecture, but in all cases 
it is advisable to minimize the amount of saved quantities. The computation of radiation reaction force 
 requires sometimes additional memory. For example, in eq. \eqref{model5} the Lorentz force from a previous
time step needs to be saved. This implies that three space components of the force are to be added for each particle in the simulation. Also, the LL model requires to save electric and magnetic fields from the previous timestep, which results in six extra quantities per grid point. In addition to adding memory requirements (more RAM needed for the problem of the same size), this would also double the amount of communications related to the exchange of field values near the boundary between the neighbouring nodes.

The additional computational cost is summarized in Table 2. The models that do not require any additional memory  (B08, S09 and LLR)  also have low overhead in terms of flop. 

Therefore, any of these three models can be introduced without affecting significantly the performance of a PIC code. We chose the LLR model and implemented it in OSIRIS pusher with the fourth-order Runge Kutta integration method. 

\section{Conclusions}
We have compared different approaches to include classical radiation reaction in a PIC code. Out of the numerous models  available in the literature, we chose a subset compatible with the standard PIC algorithm and compared their performance in modelling the electron motion in physical scenarios of interest. 

The first test was on the electron motion in a constant B-field (i.e. synchrotron), where all the models gave the same results, with a minimal departure of B08 at extremely high fields ($B \simeq 0.1~ B_C$). Interaction of electrons with a counter-propagating laser pulse (both for circular and linear polarisation) also did not show any appreciable distinction between the models. This is not surprising, because for relativistic electrons the leading order term in the radiation reaction force associated with the transverse acceleration is the same for all the compared models. However, differences arise in the presence of a longitudinal electric field with a spatio-temporal gradient. For such fields, the model F93 predicts the same recoil as the LL model, that depends on the electric field gradient. The models S09 and H08 depart from the LL results, but do predict an energy loss that depends solely on the field magnitude (not the gradient). The models B08 and LLR do not account for any energy loss in this configuration. Even though this offers a qualitative distinction between the models, the electric field needs to be on the order of the Schwinger critical field to show it, and even then the differences make for less than a percent of the total electron energy. They are, therefore, unlikely to make any real impact to the simulation results for realistic configurations. For example, these differences are negligible compared to the leading order term of the radiation reaction force for particles in the field of a laser pulse if $a_0\gamma\gg1$, which is necessarily satisfied in case of a significant radiation reaction. 

One additional question that needed to be addressed for PIC implementation is the definition of radiation reaction for a macro-particle. We note that the radiation reaction depends nonlinearly on charge and mass. To avoid overestimating the output radiation by artificially assuming coherent emission,  the real charge and mass of a single particle should be used when calculating the radiation reaction force for a macro-particle.

The models are all fit to describe the physics, but the computational cost is minimal (in terms of flop and additional memory required) for the models B08, S09 and LLR. They can be included in a massively parallel PIC code just by re-writing the particle pusher, without any additional quantities to store. We have opted to implement the LLR model in OSIRIS.   

Apart from choosing a model that accounts for radiation reaction, to correctly model the particle dynamics at extreme intensities it is essential to use the appropriate temporal resolution in the simulations. For the same laser frequency, a higher $a_0$ requires higher resolution to account correctly for the electron motion (especially if the electron is not relativistic at the start of the interaction). Detailed analysis supported by test-particle simulations was performed to facilitate the choice of temporal resolution for simulations at high intensities.

Taking all the aforementioned into account, it is possible to simulate the classical radiation reaction dominated regime with PIC codes by keeping the same parallel structure and introducing modifications only at particle push-time. Such modifications should not affect significantly the computational performance of massively parallel PIC simulations, with several models (e. g. LLR or B08) capturing the appropriate dynamics with minimal computational overhead.

\section{Acknowledgements}
We gratefully acknowledge Dr T. Grismayer and Dr S. F. Martins for fruitful discussions. This work is supported by the European Research Council (ERC-2010-AdG Grant 267841) and FCT (Portugal) grants PTDC/FIS/111720/2009, SFRH/BD/62137/2009 and SFRH/BD/39523/2007. Simulations were performed at IST cluster (Lisbon, Portugal).

\bibliographystyle{elsarticle-num}
\bibliography{Classical_RR_for_PIC.bbl}

\begin{thebibliography}{10}
\expandafter\ifx\csname url\endcsname\relax
  \def\url#1{\texttt{#1}}\fi
\expandafter\ifx\csname urlprefix\endcsname\relax\def\urlprefix{URL }\fi
\expandafter\ifx\csname href\endcsname\relax
  \def\href#1#2{#2} \def\path#1{#1}\fi

\bibitem{ELI}
\href{http://www.extreme-light-infrastructure.eu/}{The ELI Project}.
\newline\urlprefix\url{http://www.extreme-light-infrastructure.eu/}

\bibitem{model1bell}
A.~R. Bell, J.~G. Kirk, Possibility of prolific pair production with high-power
  lasers, Phys. Rev. Lett. 101~(20) (2008) 200403.
\newblock \href {http://dx.doi.org/10.1103/PhysRevLett.101.200403}
  {\path{doi:10.1103/PhysRevLett.101.200403}}.

\bibitem{Bulanov_pairsvaccuum}
S.~S. Bulanov, T.~Z. Esirkepov, A.~G.~R. Thomas, J.~K. Koga, S.~V. Bulanov,
  Schwinger limit attainability with extreme power lasers, Phys. Rev. Lett. 105
  (2010) 220407.
\newblock \href {http://dx.doi.org/10.1103/PhysRevLett.105.220407}
  {\path{doi:10.1103/PhysRevLett.105.220407}}.

\bibitem{Ridgers_solid}
C.~P. Ridgers, C.~S. Brady, R.~Duclous, J.~G. Kirk, K.~Bennett, T.~D. Arber,
  A.~P.~L. Robinson, A.~R. Bell, Dense electron-positron plasmas and
  ultraintense $\gamma${} rays from laser-irradiated solids, Phys. Rev. Lett.
  108 (2012) 165006.
\newblock \href {http://dx.doi.org/10.1103/PhysRevLett.108.165006}
  {\path{doi:10.1103/PhysRevLett.108.165006}}.

\bibitem{Elkina_rot}
N.~V. Elkina, A.~M. Fedotov, I.~Y. Kostyukov, M.~V. Legkov, N.~B. Narozhny,
  E.~N. Nerush, H.~Ruhl, Qed cascades induced by circularly polarized laser
  fields, Phys. Rev. ST Accel. Beams 14 (2011) 054401.
\newblock \href {http://dx.doi.org/10.1103/PhysRevSTAB.14.054401}
  {\path{doi:10.1103/PhysRevSTAB.14.054401}}.

\bibitem{model2noguchi}
K.~Noguchi, E.~Liang, Radiative effects on particle acceleration in
  electromagnetic dominated outflows, arXiv: astro-ph~(0412310v3) (2008)
  0412310v3.

\bibitem{model2zhidkov}
A.~Zhidkov, J.~Koga, A.~Sasaki, M.~Uesaka, Radiation damping effects on the
  interaction of ultraintense laser pulses with an overdense plasma, Phys. Rev.
  Lett. 88~(18) (2002) 185002.
\newblock \href {http://dx.doi.org/10.1103/PhysRevLett.88.185002}
  {\path{doi:10.1103/PhysRevLett.88.185002}}.

\bibitem{application1}
E.~L. Clark, K.~Krushelnick, M.~Zepf, F.~N. Beg, M.~Tatarakis, A.~Machacek,
  M.~I.~K. Santala, I.~Watts, P.~A. Norreys, A.~E. Dangor, Energetic heavy-ion
  and proton generation from ultraintense laser-plasma interactions with
  solids, Phys. Rev. Lett. 85~(8) (2000) 1654--1657.
\newblock \href {http://dx.doi.org/10.1103/PhysRevLett.85.1654}
  {\path{doi:10.1103/PhysRevLett.85.1654}}.

\bibitem{application2}
A.~Zhidkov, A.~Sasaki, T.~Tajima, Emission of mev multiple-charged ions from
  metallic foils irradiated with an ultrashort laser pulse, Phys. Rev. E 61~(3)
  (2000) R2224--R2227.
\newblock \href {http://dx.doi.org/10.1103/PhysRevE.61.R2224}
  {\path{doi:10.1103/PhysRevE.61.R2224}}.

\bibitem{ion_acc_rradd}
M.~Chen, A.~Pukhov, T.-P. Yu, Z.-M. Sheng, Radiation reaction effects on ion
  acceleration in laser foil interaction, Plasma Phys. Contr. F. 53~(1) (2011)
  014004.

\bibitem{limits}
D.~Iwanenko, I.~Pomeranchuk, On the maximal energy attainable in a betatron,
  Phys. Rev. 65 (1944) 343--343.
\newblock \href {http://dx.doi.org/10.1103/PhysRev.65.343}
  {\path{doi:10.1103/PhysRev.65.343}}.

\bibitem{rorlich_1}
F.~Rohrlich, Dynamics of a charged particle, Phys. Rev. E 77 (2008) 046609.
\newblock \href {http://dx.doi.org/10.1103/PhysRevE.77.046609}
  {\path{doi:10.1103/PhysRevE.77.046609}}.

\bibitem{Piazza_solutionLL}
A.~Di~Piazza, Exact solution of the landau-lifshitz equation in a plane wave,
  Lett. Math. Phys. 83~(3) (2008) 305--313.
\newblock \href {http://dx.doi.org/10.1007/s11005-008-0228-9}
  {\path{doi:10.1007/s11005-008-0228-9}}.

\bibitem{counter}
J.~Koga, Integration of the lorentz-dirac equation: Interaction of an intense
  laser pulse with high-energy electrons, Phys. Rev. E 70~(4) (2004) 046502.
\newblock \href {http://dx.doi.org/10.1103/PhysRevE.70.046502}
  {\path{doi:10.1103/PhysRevE.70.046502}}.

\bibitem{Self_force_Derivation}
S.~E. Gralla, A.~I. Harte, R.~M. Wald, Rigorous derivation of electromagnetic
  self-force, Phys. Rev. D 80 (2009) 024031.
\newblock \href {http://dx.doi.org/10.1103/PhysRevD.80.024031}
  {\path{doi:10.1103/PhysRevD.80.024031}}.

\bibitem{Bulanov_LLLAD}
S.~V. Bulanov, T.~Z. Esirkepov, M.~Kando, J.~K. Koga, S.~S. Bulanov,
  Lorentz-abraham-dirac versus landau-lifshitz radiation friction force in the
  ultrarelativistic electron interaction with electromagnetic wave (exact
  solutions), Phys. Rev. E 84 (2011) 056605.
\newblock \href {http://dx.doi.org/10.1103/PhysRevE.84.056605}
  {\path{doi:10.1103/PhysRevE.84.056605}}.

\bibitem{Keita_neweq}
K.~Seto, H.~Nagatomo, J.~Koga, K.~Mima, Equation of motion with radiation
  reaction in ultrarelativistic laser-electron interactions, Phys. Plasmas
  18~(12) (2011) 123101.
\newblock \href {http://dx.doi.org/http://dx.doi.org/10.1063/1.3663843}
  {\path{doi:http://dx.doi.org/10.1063/1.3663843}}.

\bibitem{spohn}
H.~Spohn, The critical manifold of the lorentz-dirac equation, Europhys. Lett.
  50~(3) (2000) 287.
\newblock \href {http://dx.doi.org/10.1209/epl/i2000-00268-x}
  {\path{doi:10.1209/epl/i2000-00268-x}}.

\bibitem{Fedotov_ponder}
A.~M. Fedotov, N.~V. Elkina, E.~G. Gelfer, N.~B. Narozhny, H.~Ruhl, Radiation
  friction versus ponderomotive effect, Phys. Rev. A 90 (2014) 053847.
\newblock \href {http://dx.doi.org/10.1103/PhysRevA.90.053847}
  {\path{doi:10.1103/PhysRevA.90.053847}}.

\bibitem{Dino_RR}
Y.~Kravets, D.~Noble, A.and~Jaroszynski, Radiation reaction effects on the
  interaction of an electron with an intense laser pulse, Phys. Rev. E 88
  (2013) 011201.
\newblock \href {http://dx.doi.org/10.1103/PhysRevE.88.011201}
  {\path{doi:10.1103/PhysRevE.88.011201}}.

\bibitem{Noble_2013_rrmodel}
A.~Noble, D.~A. Burton, J.~Gratus, D.~A. Jaroszynski, A kinetic model of
  radiating electrons, Journal of Mathematical Physics 54~(4) (2013) --.
\newblock \href {http://dx.doi.org/http://dx.doi.org/10.1063/1.4798796}
  {\path{doi:http://dx.doi.org/10.1063/1.4798796}}.

\bibitem{RRinCascade_classical}
A.~Zhidkov, S.~Masuda, S.~S. Bulanov, J.~Koga, T.~Hosokai, R.~Kodama, Radiation
  reaction effects in cascade scattering of intense, tightly focused laser
  pulses by relativistic electrons: Classical approach, Phys. Rev. ST Accel.
  Beams 17 (2014) 054001.
\newblock \href {http://dx.doi.org/10.1103/PhysRevSTAB.17.054001}
  {\path{doi:10.1103/PhysRevSTAB.17.054001}}.

\bibitem{Rohrlich_derivation}
F.~Rohrlich, The correct equation of motion of a classical point charge, Phys.
  Lett. A 283 (2001) 276 -- 278.
\newblock \href
  {http://dx.doi.org/http://dx.doi.org/10.1016/S0375-9601(01)00264-X}
  {\path{doi:http://dx.doi.org/10.1016/S0375-9601(01)00264-X}}.

\bibitem{Tamburini3d}
M.~Tamburini, T.~V. Liseykina, F.~Pegoraro, A.~Macchi,
  Radiation-pressure-dominant acceleration: Polarization and radiation reaction
  effects and energy increase in three-dimensional simulations, Phys. Rev. E 85
  (2012) 016407.
\newblock \href {http://dx.doi.org/10.1103/PhysRevE.85.016407}
  {\path{doi:10.1103/PhysRevE.85.016407}}.

\bibitem{Bulanov_chi}
S.~V. Bulanov, T.~Z. Esirkepov, Y.~Hayashi, M.~Kando, H.~Kiriyama, J.~K. Koga,
  K.~Kondo, H.~Kotaki, A.~S. Pirozhkov, S.~S. Bulanov, A.~G. Zhidkov, P.~Chen,
  D.~Neely, Y.~Kato, N.~B. Narozhny, G.~Korn, On the design of experiments for
  the study of extreme field limits in the interaction of laser with
  ultrarelativistic electron beam, Nucl. Instrum. Meth. A 660~(1) (2011) 31 --
  42.
\newblock \href
  {http://dx.doi.org/http://dx.doi.org/10.1016/j.nima.2011.09.029}
  {\path{doi:http://dx.doi.org/10.1016/j.nima.2011.09.029}}.

\bibitem{Tikhonchuk}
T.~Schlegel, V.~T. Tikhonchuk, Classical radiation effects on relativistic
  electrons in ultraintense laser fields with circular polarization, New J.
  Phys 14~(7) (2012) 073034.
\newblock \href {http://dx.doi.org/10.1088/1367-2630/14/7/073034}
  {\path{doi:10.1088/1367-2630/14/7/073034}}.

\bibitem{Bulanov_review}
S.~Bulanov, T.~Esirkepov, J.~Koga, T.~Tajima, Interaction of electromagnetic
  waves with plasma in the radiation-dominated regime, Plasma Phys. Rep. 30~(3)
  (2004) 196--213.
\newblock \href {http://dx.doi.org/10.1134/1.1687021}
  {\path{doi:10.1134/1.1687021}}.

\bibitem{ThomasRR}
A.~G.~R. Thomas, C.~P. Ridgers, S.~S. Bulanov, B.~J. Griffin, S.~P.~D. Mangles,
  Strong radiation-damping effects in a gamma-ray source generated by the
  interaction of a high-intensity laser with a wakefield-accelerated electron
  beam, Phys. Rev. X 2 (2012) 041004.
\newblock \href {http://dx.doi.org/10.1103/PhysRevX.2.041004}
  {\path{doi:10.1103/PhysRevX.2.041004}}.

\bibitem{emittance_decrease}
E.~Esarey, Laser cooling of electron beams via thomson scattering, Nucl. Instr.
  Meth. Phys. Res. 455~(1) (2000) 7 -- 14.
\newblock \href {http://dx.doi.org/10.1016/S0168-9002(00)00685-9}
  {\path{doi:10.1016/S0168-9002(00)00685-9}}.

\bibitem{Pukhov_rrtrapp}
L.~L. Ji, A.~Pukhov, I.~Y. Kostyukov, B.~F. Shen, K.~Akli, Radiation-reaction
  trapping of electrons in extreme laser fields, Phys. Rev. Lett. 112 (2014)
  145003.
\newblock \href {http://dx.doi.org/10.1103/PhysRevLett.112.145003}
  {\path{doi:10.1103/PhysRevLett.112.145003}}.

\bibitem{Harvey_inpulse}
C.~Harvey, M.~Marklund, Radiation damping in pulsed gaussian beams, Phys. Rev.
  A 85 (2012) 013412.
\newblock \href {http://dx.doi.org/10.1103/PhysRevA.85.013412}
  {\path{doi:10.1103/PhysRevA.85.013412}}.

\bibitem{Capdssus_ionwithrr}
R.~Capdessus, E.~d'Humi\`eres, V.~T. Tikhonchuk, Modeling of radiation losses
  in ultrahigh power laser-matter interaction, Phys. Rev. E 86 (2012) 036401.
\newblock \href {http://dx.doi.org/10.1103/PhysRevE.86.036401}
  {\path{doi:10.1103/PhysRevE.86.036401}}.

\bibitem{JaroschekHoshino}
C.~H. Jaroschek, M.~Hoshino, Radiation-dominated relativistic current sheets,
  Phys. Rev. Lett. 103 (2009) 075002.
\newblock \href {http://dx.doi.org/10.1103/PhysRevLett.103.075002}
  {\path{doi:10.1103/PhysRevLett.103.075002}}.

\bibitem{theory_rrinmr}
D.~A. Uzdensky, J.~C. McKinney, Magnetic reconnection with radiative cooling.
  i. optically thin regime, Phys. Plasmas 18~(4) (2011) 042105.
\newblock \href {http://dx.doi.org/http://dx.doi.org/10.1063/1.3571602}
  {\path{doi:http://dx.doi.org/10.1063/1.3571602}}.

\bibitem{Angelo_astro}
M.~D'Angelo, L.~Fedeli, A.~Sgattoni, F.~Pegoraro, A.~Macchi, Particle
  acceleration and radiation friction effects in the filamentation instability
  of pair plasmas, Mon. Not. R. Astron. Soc. 451~(4) (2015) 3460--3467.
\newblock \href {http://dx.doi.org/10.1093/mnras/stv1159}
  {\path{doi:10.1093/mnras/stv1159}}.

\bibitem{Cerruti_apj_2013}
B.~Cerutti, G.~R. Werner, D.~A. Uzdensky, M.~C. Begelman, Simulations of
  particle acceleration beyond the classical synchrotron burnoff limit in
  magnetic reconnection: An explanation of the crab flares, Astrophys. J.
  770~(2) (2013) 147.
\newblock \href {http://dx.doi.org/10.1088/0004-637X/770/2/147}
  {\path{doi:10.1088/0004-637X/770/2/147}}.

\bibitem{OSIRIS}
R.~A. Fonseca, L.~O. Silva, F.~S. Tsung, V.~K. Decyk, W.~Lu, C.~Ren, W.~B.
  Mori, S.~Deng, S.~Lee, T.~Katsouleas, J.~C. Adam, {OSIRIS: A
  three-dimensional, fully relativistic particle in cell code for modeling
  plasma based accelerators}, Vol. 2331, Springer Berlin / Heidelberg, 2002.

\bibitem{Fonseca_scaling}
R.~A. Fonseca, J.~Vieira, F.~Fiuza, A.~Davidson, F.~Tsung, W.~B. Mori, L.~O.
  Silva, Exploiting multi-scale parallelism for large scale numerical modelling
  of laser wakefield accelerators, Plasma Phys. Contr. F. 55~(12) (2013)
  124011.

\bibitem{Jackson}
J.~D. Jackson, Classical Electrodynamics Third Edition, 3rd Edition, Wiley,
  1998.

\bibitem{rorlich_LAD}
F.~Rohrlich, The dynamics of a charged sphere and the electron, Am. J. Phys.
  65~(11) (1997) 1051--1056.
\newblock \href {http://dx.doi.org/10.1119/1.18719}
  {\path{doi:10.1119/1.18719}}.

\bibitem{ArthurY}
A.~D. {Yaghjian}, {Relativistic Dynamics of a Charged Sphere: Updating the
  Lorentz-Abraham Model}, {Springer-Verlag, Berlin, Heidelberg}, 1992.

\bibitem{LAD_backwards}
F.~V. Hartemann, A.~K. Kerman, Classical theory of nonlinear compton
  scattering, Phys. Rev. Lett. 76~(4) (1996) 624--627.
\newblock \href {http://dx.doi.org/10.1103/PhysRevLett.76.624}
  {\path{doi:10.1103/PhysRevLett.76.624}}.

\bibitem{SokolovRenorm}
I.~Sokolov, Renormalization of the lorentz-abraham-dirac equation for radiation
  reaction force in classical electrodynamics, J. Exp. Theor. Phys. 109~(2)
  (2009) 207--212.
\newblock \href {http://dx.doi.org/10.1134/S1063776109080044}
  {\path{doi:10.1134/S1063776109080044}}.

\bibitem{LLbook}
L.~D. {Landau}, E.~M. {Lifshitz}, {The Classical Theory of Fields}, Butterworth
  Heinemann, 1975.

\bibitem{model3sokolov}
I.~V. Sokolov, N.~M. Naumova, J.~A. Nees, G.~A. Mourou, V.~P. Yanovsky,
  Dynamics of emitting electrons in strong laser fields, Phys. Plasmas 16~(9)
  (2009) 093115.
\newblock \href {http://dx.doi.org/10.1063/1.3236748}
  {\path{doi:10.1063/1.3236748}}.

\bibitem{Hededal}
C.~B. Hededal, {Gamma-Ray Bursts, Collisionless Shocks and Synthetic Spectra},
  Ph.D. thesis, University of Copenhagen, Danmark (2008).

\bibitem{model5ford}
G.~W. Ford, R.~F. O'Connell, Relativistic form of radiation reaction, Phys.
  Lett. A 174~(3) (1993) 182 -- 184.
\newblock \href {http://dx.doi.org/DOI: 10.1016/0375-9601(93)90755-O}
  {\path{doi:DOI: 10.1016/0375-9601(93)90755-O}}.

\bibitem{TamburiniLL}
M.~Tamburini, F.~Pegoraro, A.~DiPiazza, C.~H. Keitel, A.~Macchi, Radiation
  reaction effects on radiation pressure acceleration, New J. Phys 12~(12)
  (2010) 123005.

\bibitem{qed_class_rr}
A.~Ilderton, G.~Torgrimsson, Radiation reaction in strong field \{QED\},
  Physics Letters B 725~(4Ð5) (2013) 481 -- 486.
\newblock \href
  {http://dx.doi.org/http://dx.doi.org/10.1016/j.physletb.2013.07.045}
  {\path{doi:http://dx.doi.org/10.1016/j.physletb.2013.07.045}}.

\bibitem{QED1}
C.~Bula, K.~T. McDonald, E.~J. Prebys, C.~Bamber, S.~Boege, T.~Kotseroglou,
  A.~C. Melissinos, D.~D. Meyerhofer, W.~Ragg, D.~L. Burke, R.~C. Field,
  G.~Horton-Smith, A.~C. Odian, J.~E. Spencer, D.~Walz, S.~C. Berridge, W.~M.
  Bugg, K.~Shmakov, A.~W. Weidemann, Observation of nonlinear effects in
  compton scattering, Phys. Rev. Lett. 76 (1996) 3116--3119.
\newblock \href {http://dx.doi.org/10.1103/PhysRevLett.76.3116}
  {\path{doi:10.1103/PhysRevLett.76.3116}}.

\bibitem{QED2}
D.~L. Burke, R.~C. Field, G.~Horton-Smith, J.~E. Spencer, D.~Walz, S.~C.
  Berridge, W.~M. Bugg, K.~Shmakov, A.~W. Weidemann, C.~Bula, K.~T. McDonald,
  E.~J. Prebys, C.~Bamber, S.~J. Boege, T.~Koffas, T.~Kotseroglou, A.~C.
  Melissinos, D.~D. Meyerhofer, D.~A. Reis, W.~Ragg, Positron production in
  multiphoton light-by-light scattering, Phys. Rev. Lett. 79 (1997) 1626--1629.
\newblock \href {http://dx.doi.org/10.1103/PhysRevLett.79.1626}
  {\path{doi:10.1103/PhysRevLett.79.1626}}.

\bibitem{First_PRL}
M.~Vranic, J.~L. Martins, J.~Vieira, R.~A. Fonseca, L.~O. Silva, All-optical
  radiation reaction at $1{0}^{21}\mathrm{W}/{\mathrm{cm}}^{2}$, Phys. Rev.
  Lett. 113 (2014) 134801.
\newblock \href {http://dx.doi.org/10.1103/PhysRevLett.113.134801}
  {\path{doi:10.1103/PhysRevLett.113.134801}}.

\bibitem{Ruhl_threshold}
Y.~Hadad, L.~Labun, J.~Rafelski, N.~Elkina, C.~Klier, H.~Ruhl, Effects of
  radiation reaction in relativistic laser acceleration, Phys. Rev. D 82 (2010)
  096012.
\newblock \href {http://dx.doi.org/10.1103/PhysRevD.82.096012}
  {\path{doi:10.1103/PhysRevD.82.096012}}.

\bibitem{capturePiazza}
A.~Di~Piazza, K.~Z. Hatsagortsyan, C.~H. Keitel, Strong signatures of radiation
  reaction below the radiation-dominated regime, Phys. Rev. Lett. 102 (2009)
  254802.
\newblock \href {http://dx.doi.org/10.1103/PhysRevLett.102.254802}
  {\path{doi:10.1103/PhysRevLett.102.254802}}.

\bibitem{esarey_relmass}
E.~Esarey, S.~K. Ride, P.~Sprangle, Nonlinear thomson scattering of intense
  laser pulses from beams and plasmas, Phys. Rev. E 48 (1993) 3003--3021.
\newblock \href {http://dx.doi.org/10.1103/PhysRevE.48.3003}
  {\path{doi:10.1103/PhysRevE.48.3003}}.

\bibitem{Gibbon}
P.~{Gibbon}, {Short pulse laser interactions with matter: an introduction},
  2005.

\bibitem{SokolovLimits}
I.~V. Sokolov, J.~A. Nees, V.~P. Yanovsky, N.~M. Naumova, G.~A. Mourou,
  Emission and its back-reaction accompanying electron motion in
  relativistically strong and qed-strong pulsed laser fields, Phys. Rev. E
  81~(3) (2010) 036412.
\newblock \href {http://dx.doi.org/10.1103/PhysRevE.81.036412}
  {\path{doi:10.1103/PhysRevE.81.036412}}.

\bibitem{Alexey_subcycling}
A.~V. Arefiev, G.~E. Cochran, D.~W. Schumacher, A.~P.~L. Robinson, G.~Chen,
  Temporal resolution criterion for correctly simulating relativistic electron
  motion in a high-intensity laser field, Phys. Plasmas 22~(1) (2015) 013103.
\newblock \href {http://dx.doi.org/http://dx.doi.org/10.1063/1.4905523}
  {\path{doi:http://dx.doi.org/10.1063/1.4905523}}.

\bibitem{Moving_window}
K.-C. Tzeng, W.~B. Mori, C.~D. Decker, Anomalous absorption and scattering of
  short-pulse high-intensity lasers in underdense plasmas, Phys. Rev. Lett. 76
  (1996) 3332--3335.
\newblock \href {http://dx.doi.org/10.1103/PhysRevLett.76.3332}
  {\path{doi:10.1103/PhysRevLett.76.3332}}.

\bibitem{LWFAscailing}
W.~Lu, M.~Tzoufras, C.~Joshi, F.~S. Tsung, W.~B. Mori, J.~Vieira, R.~A.
  Fonseca, L.~O. Silva, Generating multi-gev electron bunches using single
  stage laser wakefield acceleration in a 3d nonlinear regime, Phys. Rev. ST
  Accel. Beams 10 (2007) 061301.
\newblock \href {http://dx.doi.org/10.1103/PhysRevSTAB.10.061301}
  {\path{doi:10.1103/PhysRevSTAB.10.061301}}.

\bibitem{book_simulation}
C.~K. Birdsall, A.~B. Langdon, {Plasma Physics via Computer Simulation},
  {Taylor \& Francis}, 2004.

\end{thebibliography}







\end{document}